\documentclass[twocolumn, twocolappendix]{aastex63}

\usepackage{amsmath}
\usepackage{lipsum}
\shorttitle{PeV neutrinos from blazars}
\shortauthors{Das et al.}

\begin{document}
\vspace*{-2.0cm}

\title{PeV-EeV neutrinos from gamma-ray blazars due to ultrahigh-energy cosmic-ray propagation
}

\correspondingauthor{Saikat Das}
\email{saikatdas@rri.res.in}

\author[0000-0001-5796-225X]{Saikat Das}
\affiliation{Astronomy \& Astrophysics Group, Raman Research Institute, Bengaluru 560080, Karnataka, India}

\author[0000-0002-1188-7503]{Nayantara Gupta}
\affiliation{Astronomy \& Astrophysics Group, Raman Research Institute, Bengaluru 560080, Karnataka, India}
\email{nayan@rri.res.in}

\author[0000-0002-0130-2460]{Soebur Razzaque}
\affiliation{Centre for Astro-Particle Physics (CAPP) and Department of Physics, University of Johannesburg, \\PO Box 524, Auckland Park 2006, South Africa}
\email{srazzaque@uj.ac.za}




\begin{abstract}
Blazars are potential candidates of cosmic-ray acceleration up to ultrahigh energies ($E\gtrsim10^{18}$ eV). For an efficient cosmic-ray injection from blazars, $p\gamma$ collisions with the extragalactic background light (EBL) and cosmic microwave background (CMB) can produce neutrino spectrum with peaks near PeV and EeV energies, respectively. We analyze the contribution of these neutrinos to the diffuse background measured by the IceCube neutrino observatory. The fraction of neutrino luminosity originating from individual redshift ranges is calculated using the distribution of BL Lacs and FSRQs provided in the \textit{Fermi}-LAT 4LAC catalog. Furthermore, we use a luminosity dependent density evolution to find the neutrino flux from unresolved blazars. The results obtained in our model indicate that as much as $\approx10\%$ of the flux upper bound at a few PeV energies can arise from cosmic-ray interactions on EBL. The same interactions will also produce secondary electrons and photons, initiating electromagnetic cascades. The resultant photon spectrum is limited by the isotropic diffuse $\gamma$-ray flux measured between 100 MeV and 820 GeV. The latter, together with the observed cosmic-ray flux at $E>10^{16.5}$ eV, can constrain the baryonic loading factor depending on the maximum cosmic-ray acceleration energy.
\end{abstract}


\section{Introduction} \label{sec:introduction}

The IceCube neutrino observatory in Antarctica has been detecting neutrino events between 100 GeV and a few PeV for the last ten years. The existence of a diffuse flux of astrophysical neutrinos (from $\sim 10$ TeV to a few PeV) has been established with more than $5\sigma$ significance \citep{Aartsen_2013, Aartsen_2014}, inconsistent with an atmospheric origin. The flux is isotropic across the entire sky and in all flavors, indicating the candidate sources are extragalactic. Although, a non-negligible Galactic component may also be present \citep{Razzaque_2013, Ahlers:2013xia, Taylor_2014, Neronov_2014, Aartsen_2017c}. The observed all-flavor spectrum can be explained by a single power-law with a best-fit spectral index of $2.89^{+0.2}_{-0.19}$ \citep{Aartsen_2015a, Schneider_2019}. While $\gamma$-rays can be produced in both leptonic and hadronic processes, neutrinos are an exclusive probe of hadronic interactions. They point back to their sources and are important messengers of cosmic-ray acceleration. 
IceCube has a real-time alert program that selects high-energy muon neutrino events ($\gtrsim 100$ TeV) for rapid detection of electromagnetic counterparts arriving from the same direction \citep{Aartsen_2017d}. The IceCube-170922A alert led to the $3\sigma$ association of a flaring $\gamma$-ray blazar TXS 0506+056 in spatial and temporal coincidence with a $\sim0.3$ PeV muon track \citep{Aartsen_2018a, Aartsen_2018b}. Other less significant candidate events with blazar-neutrino spatial coincidence have also been identified thereafter \citep{Steeghs_2019, Garrappa_2019, Franckowiak_2020}. 

Blazars are a subclass of radio-loud AGNs, which have their jet emission closely aligned along the observer's line-of-sight. Blazars are further classified into FSRQs and BL Lacs depending on the prominence of emission lines in the observed spectral energy distribution (SED). They have been long considered as the sources of high-energy neutrinos \citep{Eichler_1979, Sikora_1987, Berezinskii_1989, Stecker_1991, Mannheim_1992, Szabo_1994, Atoyan_2001, Becker_2008, Murase_2014, Petropoulou_2015, Palladino_2019, Yuan_2020, Rodrigues:2020pli}. Once accelerated inside the jet, the electrons and positrons lose energy via synchrotron emission and inverse-Compton (IC) scattering. Cosmic rays can interact with ambient matter ($pp$ collision) or radiation ($p\gamma$ process) to produce charged and neutral pions, that decays to produce $\gamma$ rays and neutrinos. The target photons for $p\gamma$ process are provided by the synchrotron/IC photons, or external photons from the broad-line region (BLR), accretion disk (AD), or dusty torus (DT). Despite plausible association of blazars with neutrinos observed in a few cases, it is difficult to explain the observed neutrino spectrum by blazars alone. Using the \textit{Fermi}-LAT 2LAC catalog to define search positions, IceCube collaboration has shown that astrophysical neutrinos from blazars can only account for $\sim7-27\%$ of the observed flux between 10 TeV and 2 PeV. The limit depends on various weighting schemes accounting for the relative neutrino flux from a specific source \citep{Aartsen_2017b}. Thus, the origin of these neutrinos is still controversial. 

Cosmic rays can also escape their sources without undergoing $p\gamma$ interactions inside. Ultrahigh-energy cosmic rays (UHECRs; $E\gtrsim10^{18}$ eV) with energies greater than the threshold of photopion production on CMB ($E_{\rm th}^{p,\pi}\approx6\cdot 10^{19}$ eV) can yield EeV neutrinos.
In this work, we investigate the neutrino flux produced by cosmic rays that escape from the blazar emission region and interact with the EBL, comprising of UV/optical/IR photons, as well as the radio photons of the CMB. The threshold for photopion production on EBL peaks at $E_{\rm th}^{p,\pi}\approx10^{17}$ eV, resulting in most neutrino events to lie between a few PeV to tens of PeV. The exact value of the peak energy and the width of the spectrum depends on the injection spectrum and $E_{p,\rm max}$. It has been shown, using AGN source density evolution derived from x-ray luminosities, that neutrinos resulting from such a process can account for the PeV neutrino flux measured by IceCube \citep{Kalashev_2013}. An angular correlation of these neutrino events with the blazar population is difficult to obtain, since cosmic rays are deflected from the source direction by extragalactic magnetic field (EGMF). The lepto-hadronic SED modeling of blazars reveals that protons can be accelerated up to ultrahigh energies in the comoving jet frame. Their escape depends on the seed photon density of $p\gamma$ interactions inside the AGN jet \citep{Mastichiadis_1996, Murase_2012, Razzaque_2012, Bottcher_2013, Tavecchio_2014b, Xue_2019, Sahu_2019, Das_2020}.

The 8-yr \textit{Fermi}-LAT 4LAC catalog contains $80\%$ more sources compared to the previous 3LAC catalog \citep{Ackermann_2015a, Ajello_2020}. The following analysis investigates the cumulative neutrino spectrum from 4LAC blazars, resulting in the aforementioned process. Furthermore, we use the known luminosity function of BL Lacs and FSRQs to extrapolate the sources below the \textit{Fermi}-LAT sensitivity and include the contribution from these unresolved sources \citep{Ajello_2012, Ajello_2014}. The injected cosmic-ray luminosity is assumed to be proportional to the observed point-source $\gamma$-ray luminosity by a constant factor. The electrons and photons, simultaneously produced with neutrinos, undergo electromagnetic cascade to produce $\gamma$-rays. Its contribution to the isotropic diffuse $\gamma$-ray background (IGRB), measured by \textit{Fermi}-LAT between 100 MeV and 820 GeV, is also calculated \citep{Ackermann_2015b}. We do not prefer any cosmic-ray acceleration mechanism over others proposed hitherto for this scenario. In Sec.~\ref{sec:Results}, we summarize  the basic methodology of neutrino flux calculations and present our results for various source parameters. We discuss the implications of our work in Sec.~\ref{sec:discussions} and draw the conclusions in Sec.~\ref{sec:conclusions}.


\section{Results} \label{sec:Results}

\subsection{Cosmic-ray injection and propagation}
The integrated $\gamma$-ray flux, between 100 MeV and 100 GeV, observed from the direction of resolved blazars is reported in the 4LAC catalog by the quantity $F_{100}$ in units of erg cm$^{-2}$ s$^{-1}$. We represent the K-corrected $\gamma$-ray luminosity values, corresponding to this flux and the redshift of the sources, by $L_{100}$. Thus we have
\begin{align}
F_{100}&= \int_{100 \ \text{MeV}}^{100 \ \text{GeV}} \epsilon_{\gamma}\dfrac{dN}{d\epsilon_\gamma} d\epsilon_\gamma \\
L_{100}&= 4\pi d_L^2 F_{100} (1+z)^{\Gamma - 2} 
\end{align}
where $z$ is the redshift of the source, $d_L$ is the luminosity distance, and $\Gamma$ is the slope of the observed $\gamma$-ray spectrum. The $\gamma$-ray luminosity inside the source $L'_{100}$ is usually smaller than the observed luminosity $L_{100}$ due to relativistic beaming. The intrinsic $\gamma$-ray luminosity in the comoving frame of the jet is Doppler boosted by the factor $L_{100}=(\delta_e^6/\Gamma_e^2)L'_{100}$ for FSRQs, and $L_{100}=\delta_e^4L'_{100}$ for BL Lacerate objects \citep{Dermer&Menon}. Here $\delta_e$ and $\Gamma_e$ are the doppler factor and bulk Lorentz factor of the emitting region in the relativistic jet. We define the baryonic loading factor $\eta$ to be a constant, for all blazars, that connects the intrinsic kinetic power $L'_p$ in cosmic rays with the intrinsic $\gamma$-ray luminosity $L'_{100}$.
\begin{equation}
L'_p = \eta L'_{100} 
\end{equation}

The emission region of the jet contains both leptons and hadrons. We assume the observed $\gamma$-ray flux $F_{100}$ originates from only leptonic processes inside the source, and baryons carry much more energy than leptons. For the analysis presented in this paper, we consider only protons with $E>10$ PeV are injected as cosmic rays.
Now, the cosmic-ray luminosity outside the jet (AGN frame) transforms as $L_p = \Gamma_e^2 L'_p$ \citep{Celotti_2007}. Hence, in the observer frame, the scaling between the injected cosmic-ray luminosity ($L_p$) and the observed $\gamma$-ray luminosity ($L_{100}$) turns out to be
\begin{align}
L_p &= \Gamma_e^2 L'_p = \Gamma_e^2 \eta L'_{100} \nonumber \\
&\simeq \eta L_{100}/\Gamma_e^2 = \eta_{\rm eff} L_{100} \label{eqn:L_p}
\end{align}
where we assume $\delta_e\simeq \Gamma_e$, for jet opening angles $\theta_j\sim1/\Gamma_e$ and $\eta_{\rm eff} = \eta/\Gamma_e^2$ is the effective baryonic loading.

Once injected into the extragalactic space, the cosmic rays propagate and undergo $p\gamma$ interactions with EBL and CMB photons to produce charged and neutral pions ($p+\gamma_{\rm bg}\rightarrow p\pi^0, \text{ or } n\pi^+$). The decay of $\pi^+$ and $\pi^0$ results in neutrinos and $\gamma$-rays, respectively. Including Bethe-Heitler pair production interactions ($p+\gamma_{\rm bg}\rightarrow\mathrm{e^+e^-}$), the secondary $\mathrm{e^\pm}$ and $\gamma$-rays initiate electromagnetic (EM) cascades down to GeV energies. The high-energy photons undergo pair-production processes, while the $\mathrm{e^\pm}$ can undergo triplet pair-production, synchrotron, and also IC process, upscattering the background photons to higher energies. The resulting photon spectrum peaks at $\sim$TeV energies, thus contributing to the IGRB flux measured by \textit{Fermi}-LAT \citep{Ackermann_2015b}. The neutrinos propagate unhindered by interactions and undeflected by cosmic magnetic fields to reach the observer and contribute to the isotropic diffuse neutrino background at PeV energies.

Cosmic rays injected from the blazars are propagated using the \textsc{CRPropa 3} simulation framework to obtain the neutrino, $\gamma$-ray and cosmic ray flux at Earth \citep{Batista_16}. We consider an injection spectrum of the shape $dN/dE \propto E^{-\alpha_p}$. \textsc{CRPropa 3} allows us to include all energy loss processes and also takes into account the adiabatic expansion of the universe. Since we are interested in the diffuse fluxes, a null magnetic field is considered for CR propagation. The propagation of secondary EM particles, initiating the electromagnetic cascade, is solved using the \textsc{DINT} code \citep{Lee_98, Heiter_18}. The EM cascade of secondary $\mathrm{e^\pm}$ and $\gamma$ photons depends on the pervading magnetic field, and we set the rms field strength $B_{\rm rms}=0.1$ nG for the EGMF. We use the \cite{Gilmore_2012} EBL model for both cosmic ray interactions and EM cascade.

\begin{table*}[htp]
\caption{\label{tab:LDDE} Parameter values for the best-fit LDDE model}
 \begin{ruledtabular}
 \begin{tabular}{lllllllllllll}
 Sample & $A$ & $L_*$/$10^{48}$ & $\gamma_1$ & $\gamma_2$ & $z_c^*$ & $\alpha$ & $p_1^*$ & $p_2$ & $\mu$ & $\sigma$ \\
 & [Gpc$^{-3}$] & [erg s$^{-1}$] & & & & & & & & & & \\
 \hline
 BL Lac & $3.39^{+7.44}_{-2.13}$ & $0.28^{+0.43}_{-0.21}$ & $0.27^{+0.26}_{-0.46}$ & $1.86^{+0.86}_{-0.48}$ & $1.34_{-0.27}^{+0.22}$ & $0.0453^{+0.0498}_{-0.0652}$ & 2.24$_{+1.25}^{-1.07}$ & $-7.37_{-5.43}^{+2.95}$ & 2.10$^{+0.03}_{-0.03}$ & 0.26$^{+0.02}_{-0.02}$ \\
 FSRQ & $3.06^{+0.23}_{-0.23}$ & $0.84^{+0.49}_{-0.49}$ & $0.21^{+0.12}_{-0.12}$ & $1.58^{+0.27}_{-0.27}$ & $1.47_{-0.16}^{+0.16}$ & $0.21^{+0.03}_{-0.03}$ & 7.35$_{+1.74}^{-1.74}$ & $-6.51_{-1.97}^{+1.97}$ & 2.44$^{+0.01}_{-0.01}$ & 0.18$^{+0.01}_{-0.01}$
 \end{tabular}
 \end{ruledtabular}
\end{table*}

The efficiency of neutrino production by cosmic rays will depend on the number of interaction lengths, and hence, on the redshift of their sources. We assume cosmic rays are injected from 10 PeV up to a maximum energy $E_{p, \rm max}$. The fraction of injected cosmic-ray energy ($\mathcal{E}_p$), from a redshift $z$, carried away by cascade photons ($\mathcal{E}_\gamma$) and secondary neutrinos ($\mathcal{E}_\nu$) are given by $f_\nu(z)$ and $f_\gamma(z)$, respectively. 
\begin{align}
f_{\nu} = \dfrac{\mathcal{E}_\nu(z)}{\mathcal{E}_p} = \dfrac{1}{\mathcal{E}_p} \times \int_{10 \text{\ TeV}}^{E_{p, \rm max}} \epsilon_{\nu}(dN/d\epsilon_{\nu}) d\epsilon_{\nu} \label{eqn:f_neu}\\
f_{\gamma} = \dfrac{\mathcal{E}_\gamma(z)}{\mathcal{E}_p} = \dfrac{1}{\mathcal{E}_p} \times \int_{10 \text{\ MeV}}^{E_{p, \rm max}} \epsilon_{\gamma}(dN/d\epsilon_{\gamma}) d\epsilon_{\gamma} \label{eqn:f_gamma}
\end{align}
All quantities in the above equations are calculated in the observer frame,
using one-dimensional simulations in \textsc{CRPropa} 3, for a fixed $\alpha_p$ and different source distances. The value of $\mathcal{E}_p$ is fixed at all redshifts. These quantities signify the energy loss fraction of protons in various secondary channels, and hence a null intergalactic magnetic field is assumed. The latter can eventually deflect the parent cosmic rays, smeared over a solid angle $\Omega$, thus resulting in a diffuse secondary flux.
The normalization to the neutrino luminosity from a blazar at redshift $z$ and $\gamma$-ray luminosity $L_{100}$ is thus obtained by the following condition (using Eqn.~\ref{eqn:L_p})
\begin{align}
L_\nu^{\rm obs} = f_\nu L_p = f_\nu \eta_{\rm eff} L_{100} \label{eqn:L_neu}
\end{align}
%
The same expression also holds for secondary photon luminosity $L_{\gamma}^{\rm obs}$, with $f_\nu$ replaced by $f_\gamma$. Summing over all sources at all redshifts and different directions, the cumulative diffuse neutrino spectrum at Earth is
\begin{equation}
F_\nu^{\rm tot} = \dfrac{1}{\Omega}\sum_{i} \bigg(\dfrac{L_\nu^{\rm obs}}{\Omega d_L^2}\bigg)_i \ \ \text{[erg cm$^{-2}$ s$^{-1}$ sr$^{-1}$]} \label{eqn:F_neu}
\end{equation} 
where $d_{L}$ is the luminosity distance corresponding to redshift $z$. The summation $i$ runs over the number of blazars in the sample space. We consider a conservative upper limit $\Omega=4\pi$, that gives the minimum flux possible in this scenario. 

\subsection{Blazar luminosity distribution}

The fourth catalog of \textit{Fermi}-LAT AGNs, detected over the period between August 2008 to August 2016 at high Galactic latitudes $| b|>10^\circ$, contains 2863 objects in the energy range between 50 MeV and 1 TeV. This paper uses the sources from the latest 4LAC catalog to calculate the cumulative neutrino and cascade gamma-ray fluxes from blazars, originating in cosmic-ray interactions during extragalactic propagation. There are a total of 655 FSRQs and 1067 BL Lacs listed in the entire catalog. The redshift information is available for all the FSRQs but lacks for 36\% of the BL Lacs. 
Besides, there are 1077 blazar candidates of unknown types (BCU). The number density of blazars depends on both luminosity and redshift distributions. The luminosity function (LF) is modeled as a double power-law multiplied by the photon index evolution. We use the parametrization by \cite{Ajello_2012, Ajello_2014} to evaluate the distribution of BL Lac objects and FSRQs, including unresolved sources. At $z=0$, the number of sources $N$ per comoving volume $V_c$, emitted luminosity $L_{100}$ between $0.1-100$ GeV, and slope of $\gamma$-ray flux $\Gamma$ is
\begin{align}
\Phi &(L_{100},z=0,\Gamma) = \dfrac{dN}{dL_{100} dV_cd\Gamma} = \dfrac{A}{\ln(10)L_{100}} \nonumber \\
& \times \bigg[\bigg(\dfrac{L_{100}}{L_*}\bigg)^{\gamma_1} + \bigg(\dfrac{L_{100}}{L_*}\bigg)^{\gamma_2} \bigg]^{-1} g(\Gamma, L_{100})
\end{align}

The photon index distribution $g(\Gamma, L_{100})$ is considered to be a Gaussian with the mean and dispersion given by $\mu$ and $\sigma$ respectively as,
\begin{equation}
g(\Gamma, L_{100}) = \exp{\bigg[-\dfrac{[\Gamma-\mu(L_{100})]^2}{2\sigma^2}\bigg]}
\end{equation} The mean is parametrized as a function of luminosity,
\begin{equation}
\mu(L_{100}) = \mu^* + \beta\times[\log(L_{100}) - 46]
\end{equation}

The redshift evolution is incorporated by the factor $e(z,L_{100})$, such that the luminosity dependent density evolution (LDDE) is represented as
\begin{equation}
\Phi(L_{100},z,\Gamma) = \Phi(L_{100},z=0,\Gamma) \times e(z,L_{100}) \label{eqn:num_den}
\end{equation}
The evolutionary factor is expanded as
\begin{align}
e(z,L_{100}) = & \bigg[\bigg(\dfrac{1+z}{1+z_c(L_{100})}\bigg)^{-p_1(L_{100})} \nonumber \\
& + \bigg(\dfrac{1+z}{1+z_c(L_{100})}\bigg)^{-p_2}\bigg]^{-1}
\end{align}
with the following parametrizations,
\begin{align}
z_c(L_{100}) &= z_c^* (L_{100}/10^{48})^\alpha \\
p_1(L_{100}) &= p_1^* + \tau\times[\log(L_{100})-46]
\end{align}
%
where $z_c$ is the redshift where the evolution changes sign from positive to negative, and $z_c^*$ is the redshift peak for a luminosity of $10^{48}$ erg s$^{-1}$. We use the values of the 12 parameters ($A$, $\gamma_1$, $\gamma_2$, $L_*$, $z_c^*$, $\alpha$, $p_1^*$, $p2$, $\mu$, $\sigma$, $\beta$, $\tau$) as obtained for the best-fit LDDE model, reported in \cite{Ajello_2012, Ajello_2014}. The wrong positive sign of $p_1$ and $p_2$ therein has been corrected in \cite{Ajello_2015}. We list the values in Table~\ref{tab:LDDE}. For BL Lacs, $\beta=0.0646^{+0.0234}_{-0.0207}$ and $\tau=4.92^{+1.45}_{-2.12}$, and for FSRQs, they are zero. Integrating $\Phi(L_{100}, z, \Gamma)$ gives the total number of sources, resolved and unresolved combined together. We follow the method of \cite{Palladino_2019} (see Appendix B there) and write the number density in terms of redshift $z$ and $\ell=\log_{10}(L_{100}/\text{erg s$^{-1}$})$,
\begin{equation}
\dfrac{dN}{dzd\ell d\Gamma} = \dfrac{dV_c}{dz} \times \dfrac{dL_{100}}{d\ell} \times \dfrac{dN}{dL_{100} dV_c d\Gamma} \label{eqn:jacobian}
\end{equation}

\begin{figure}
\centering
\includegraphics[width = 0.49\textwidth]{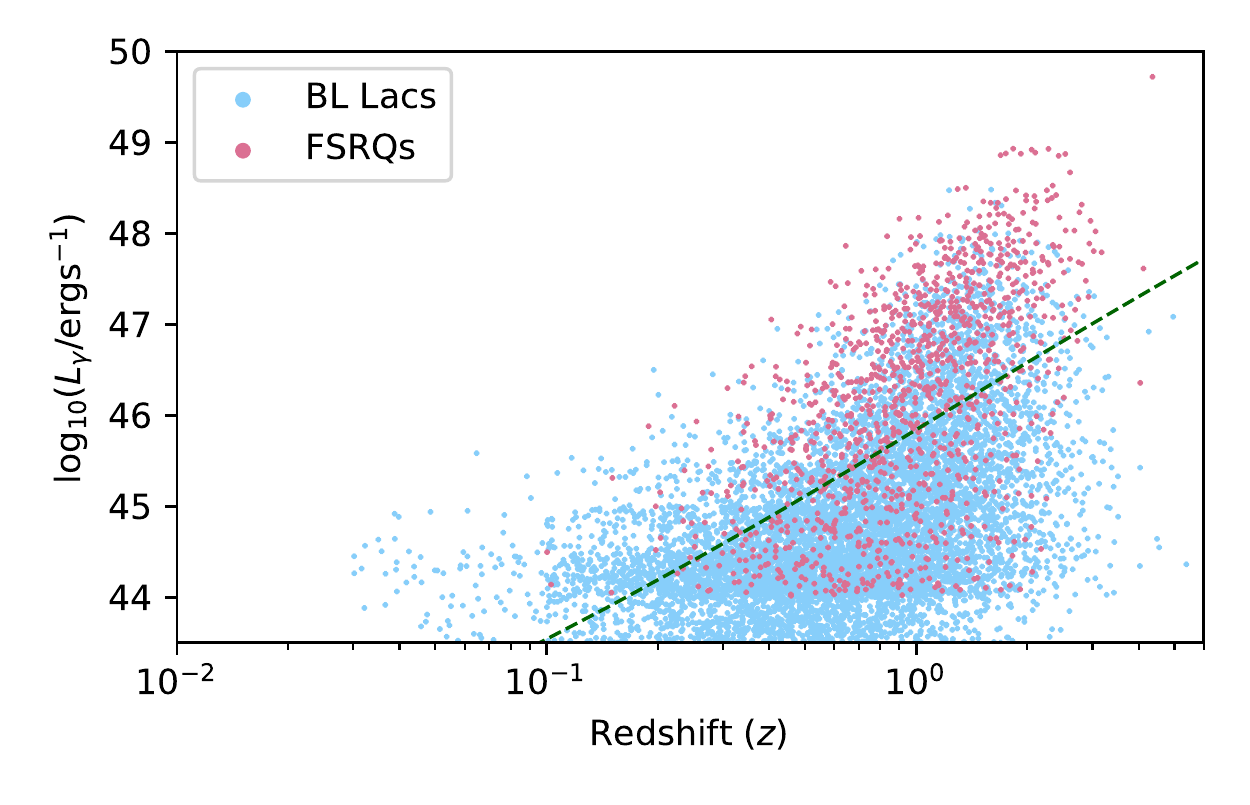}
\caption{\small{
The distribution of blazars in luminosity-redshift space according to the luminosity function deduced in \cite{Ajello_2012, Ajello_2014}. The dashed line separates the region into resolved and unresolved sources in \textit{Fermi}-LAT survey.}}
\label{fig:Lz_dist}
\end{figure}

\begin{figure*}
\centering
\includegraphics[width = 0.49\textwidth]{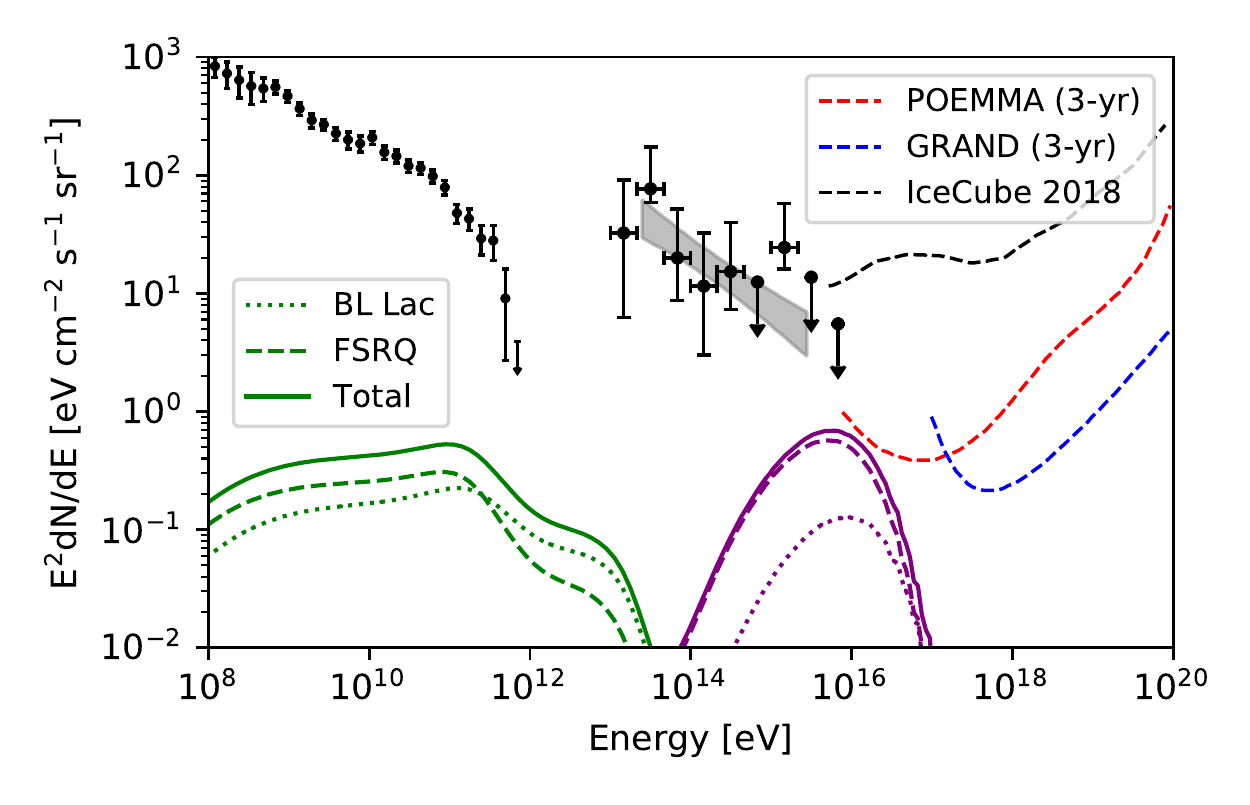}
\includegraphics[width = 0.49\textwidth]{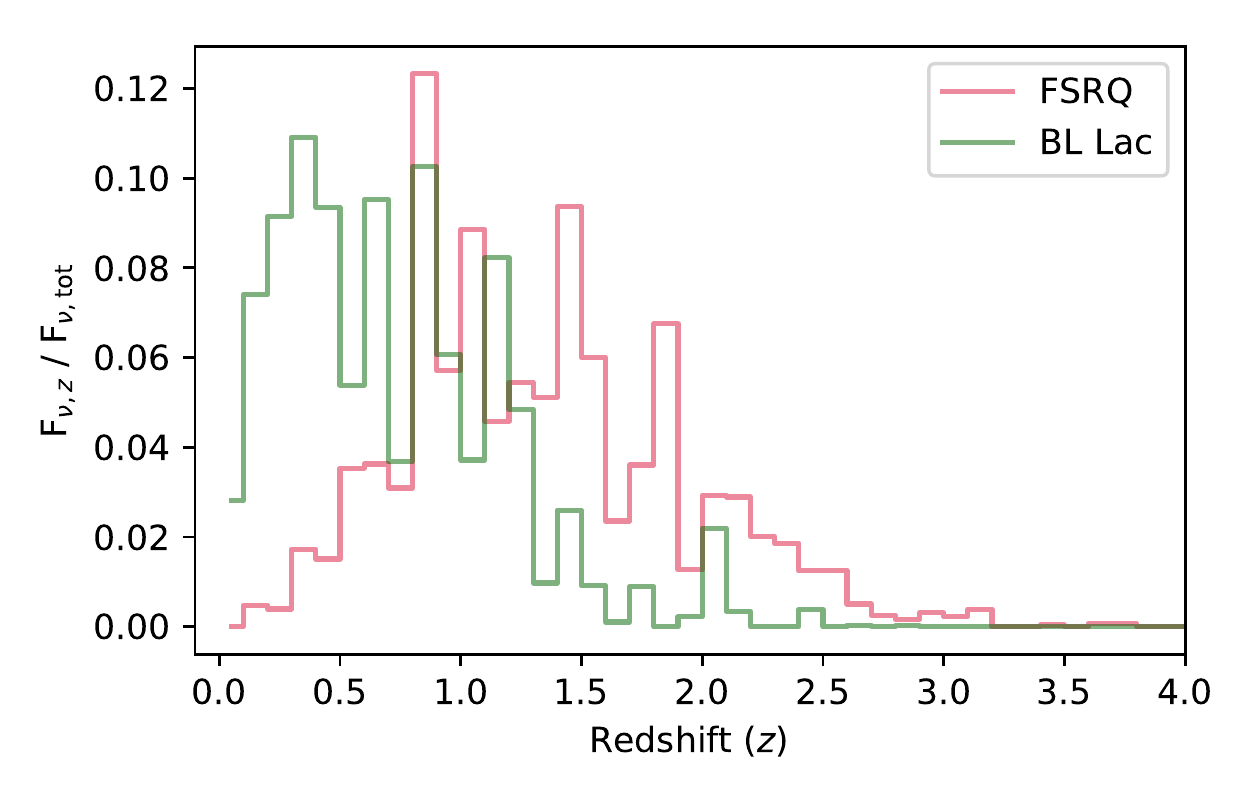}
\caption{\small{\textit{Left:} The neutrino and IGRB flux from \textit{Fermi}-detected blazars for $\eta_{\rm eff}=10.0$, and  $E_{p, \rm max}=1$ EeV. \textit{Right:} Fractional contribution to the neutrino flux from each redshift bin, relative to the individual flux from BL Lacs and FSRQs.}}
\label{fig:resolved}
\end{figure*}

To check the flux from a complete sample, we integrate Eqn.~\ref{eqn:num_den} over suitable ranges of luminosity, spectral index, and redshift. For FSRQs, $\ell\in [44.0, \ 52.0]$, $\Gamma\in [1.8, \ 3.0]$, and $z\in [0.01, \ 6.0]$. The range of values for BL Lacs are $\ell\in[43.85, \ 52]$, $\Gamma\in [1.45, \ 2.80]$, and $z\in [0.03, \ 6.0]$. A total of 9172 blazars are obtained by integrating over the entire parameter range in Eqn.~\ref{eqn:jacobian}. A representative distribution of blazars in the $\ell-z$ space is shown in the Fig.~\ref{fig:Lz_dist}. The dashed line corresponds to a flux of $\phi_\gamma=1.25\times10^{-12}$ erg cm$^{-2}$ s$^{-1}$ and $\Gamma=2$, roughly separating the region into resolved and unresolved sources above and below, respectively. This threshold flux is chosen to match the 4LAC statistics of $\sim2800$ observed blazars, including 1077 blazars of unknown type. Thus, Fig.~\ref{fig:Lz_dist} corresponds to 2072 resolved, and 5931 unresolved BL Lac objects, while there are 742 resolved and 427 unresolved FSRQs. The low-luminosity BL Lac objects ($L_{100}<10^{44}$ erg s$^{-1}$) show a negative redshift evolution and are mostly confined at low redshifts \citep{Ajello_2014}. High luminosities and redshifts are dominated by FSRQs, as expected.

\subsection{Secondary neutrino and $\gamma$-ray flux}

The maximum acceleration energy of a blazar $E_{p, \rm max}$ is determined by the escape timescale ($t_{\rm esc}$), acceleration timescale ($t_{\rm acc}$) and photohadronic interaction timescales ($t_{p\gamma}$) inside the jet. After fitting the synchrotron and IC peak by leptonic component, the SED modeling of representative BL Lac objects shows that the maximum acceleration energy of protons can extend up to $10^{19}$ eV \citep{Murase_2012, Razzaque_2012, Bottcher_2013, Xue_2019, Sahu_2019, Das_2020}. For FSRQs, we have used the values of $\delta_e$, $\Gamma_e$, and seed photon density as obtained in the modeling of CTA 102 \citep{Prince_2018}, to find that acceleration dominates up to a few times $10^{19}$ eV. This is also shown for a more generic class of quaser-hosted blazars in \cite{Murase_2014}. In the following analysis, we consider optimistic values of $E_{p, \rm max}$ based on these results. 

\begin{figure*}
\centering
\includegraphics[width = 0.49\textwidth]{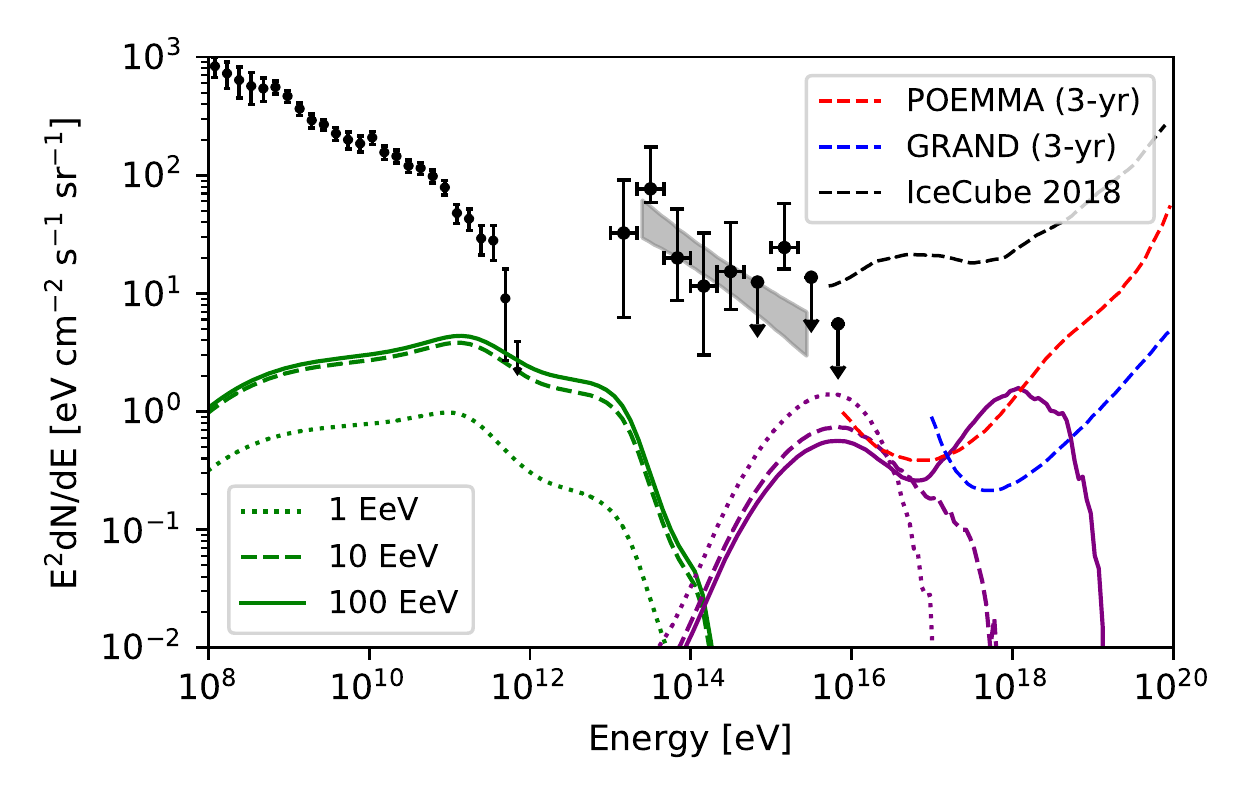}
\includegraphics[width = 0.49\textwidth]{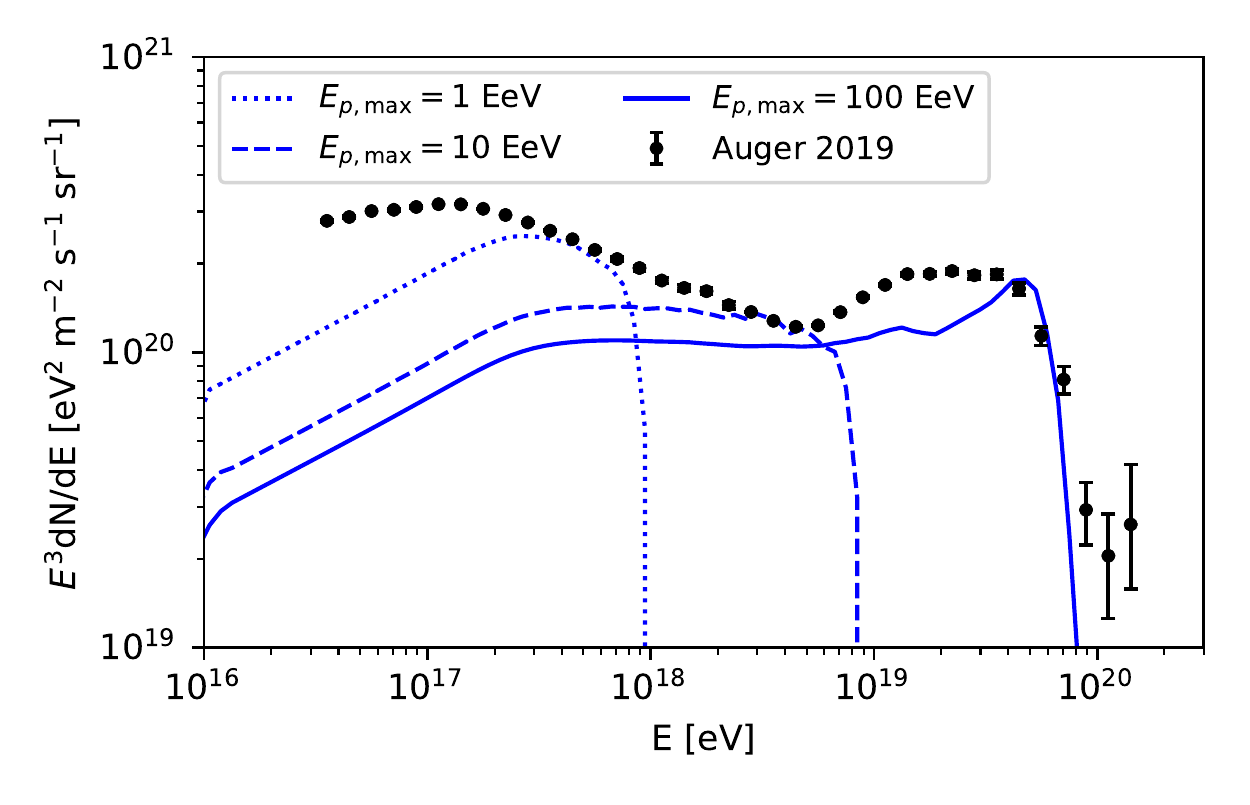}
\caption{\small{\textit{Left}: The neutrino and IGRB flux, including the unresolved blazars, for the maximum values of $\eta_{\rm eff}$ corresponding to $E_{p, \rm max}=1$, 10, and 100 EeV. \textit{Right}: The observed cosmic-ray spectrum at Earth for the maximum values of $\eta_{\rm eff}$ corresponding to $E_{p, \rm max}=1$, 10, and 100 EeV, such that the UHECR flux is not violated.}}
\label{fig:unresolved}
\end{figure*}

\subsubsection{Contributions from the resolved $\gamma$-ray blazars}

We first calculate the $\gamma$-ray and neutrino flux from \textit{Fermi}-LAT resolved sources (BL Lacs $+$ FSRQs). For simplicity, we assume the sources inject only protons as cosmic rays between 10 PeV -- 1 EeV, and fix the injection spectral index to $\alpha_{p}=2.6$. The effects of the variation of these parameters will be discussed later. For this case, we consider the value of baryonic loading factor $\eta_{\rm eff}=10$. The sources lacking redshift information are excluded from the analysis. This leaves out 381 BL Lac objects from the entire catalog. The sources are binned in a two-dimensional ($\ell$, $z$) grid, where $\ell=\log_{10}(L_{100}/\text{erg s}^{-1})$. We take the step sizes $\Delta\ell=0.5$ over the range 43.5 to 52.0, and $\Delta z=0.1$ over the range 0.0 to 6.0 that covers the entire 4LAC catalog. The mean values $\ell_m$ and $z_m$ for a given bin is used to obtain the secondary neutrino and $\gamma$-ray flux according to Eqn.~\ref{eqn:F_neu}. The number of sources $w$ in a grid provides the weight factor to the normalization of secondary fluxes. 

The resulting $\gamma$-ray and neutrino fluxes are shown in the left panel of Fig.~\ref{fig:resolved}. The neutrino spectrum peaks at an energy $E_\nu\approx6$ PeV. At around 6.3 PeV, $\overline{\nu}_e$ have increased interaction probability with ice to produce the on-shell $W^-$ boson due to Glashow-resonance (GR). However, it is challenging to detect the GR signal \citep{Biehl:2016psj, Huang:2019hgs}. The enhancement in the number of events due to the GR is only a factor of a few in the relevant energy range, as predicted by \cite{Bhattacharya_2011}, who also proposed to use it as a discriminator between the $p\gamma$ and $pp$ origin of neutrinos. Based on the current event statistics and considering no GR neutrinos are detected, this also indicates that the IceCube high-energy neutrinos originate dominantly in $p\gamma$ rather than $pp$ processes \citep{Sahu:2016qet}.  

Our calculated flux is an order of magnitude lower than the IceCube upper limit at this energy. The peak value is also comparable to the projected 3-yr sensitivity of the POEMMA detector. The contribution to this neutrino flux from each redshift bin will depend on the number of blazars and their luminosity values in that bin. The right panel of Fig.~\ref{fig:resolved} shows the fraction of neutrino luminosity coming from each redshift bin, individually for BL Lacs and FSRQs. The flux contribution from BL Lacs is approximately constant up to $z=1$ and then falls off sharply for higher $z$. Whereas the emission from FSRQs shows a peak near a redshift value of $z=1$.

\subsubsection{Contributions from the resolved and unresolved $\gamma$-ray blazars}
We show the secondary fluxes, corresponding to the distribution obtained from the luminosity function, in the left panel of Fig.~\ref{fig:unresolved}. The number of blazars $w$ in each of the ($\ell$, $z$) grid is calculated for the same values of $\Delta\ell$ and $\Delta z$ used in the preceeding case. The dotted, dashed and solid curves indicate the fluxes for the maximum allowed values of $\eta_{\rm eff}$ corresponding to each value of $E_{p, \rm max}$. The latter is derived from the UHECR flux measured by the Pierre Auger Observatory (PAO) \citep{PAO_2019}, which puts an upper bound of $\eta_{\rm eff}=11.1$, 5.8, and 4.4, for $E_{p, \rm max}=1$, 10, and 100 EeV respectively. This is shown in the right panel of Fig.~\ref{fig:unresolved}. We see that cosmic ray interactions can explain a little more than $10\%$ of the IceCube flux upper limit at $\sim$6 PeV. 
For $E_{p, \rm max}=1$ and 10 EeV, POEMMA should be able to constrain the fluxes after a few years of observation. An increase in the value of $E_{p, \rm max}$ to 10 EeV increases the cascade photon flux and saturates the IGRB background at TeV energies. 
The neutrino spectrum is broadened for higher $E_{p, \rm max}$ due to neutrinos arising from photopion interactions with the high-energy tail of the CMB spectrum. With further increase in $E_{p, \rm max}$ to 100 EeV, the GZK neutrinos becomes more prominent, and the neutrino spectrum attains a a double-humped shape, characteristic of cosmogenic neutrinos. 

The \textit{Fermi}-LAT IGRB intensity is supposed to decrease as fainter sources are resolved with future deep surveys. But, the component that we deduce here is purely diffuse, if the $\gamma$-rays from cosmic-ray interactions are not produced along the blazar line-of-sight. To maintain the constraints put by IGRB measurements, the baryonic loading factor $\eta$ must be decreased for $E_{p, \rm max}\gtrsim10^{19}$ eV. However, that in turn further decreases the neutrino flux at a few PeV energies. The maximum luminosity in the 4LAC catalog occurs for an FSRQ with $\ell=48.8$ and $z=2.534$. The value of $\eta_{\rm eff} = 11.1$ obtained for $E_{p,\rm max} = 1$ EeV and bounded by the UHECR data,  corresponds to $L_p\sim6\cdot 10^{49}$ erg s$^{-1}$ for the most luminous object. The neutrino flux obtained from an individual source, $F_\nu\propto \eta_{\rm eff} \propto \eta/\delta_e^2$, for a given value of $L_{100}$ and $E_{p, \rm max}$. To obtain the same neutrino luminosity for a lower value of $\eta$, the value of $\delta_e$ must also decrease. Indeed the value of doppler factor may vary for individual AGNs depending on the accretion rate of the central black hole. It is however not possible to extract the Doppler factor and the Lorentz factor of all the individual blazars used in this study from observations. Hence, we make the simplifying assumption $\delta_e\simeq\Gamma_e$ for the special case of $\theta\sim1/\Gamma_e$. The maximum value of the Doppler factor can be 2$\Gamma_e$, in that case the required luminosity in injected protons ($L_p$) is four times lower.


\section{Discussions} \label{sec:discussions}

In this work, we apply a multi-messenger approach to constrain the diffuse flux of PeV-EeV neutrinos originating from cosmic-ray interactions on EBL and CMB. We consider blazars as the candidate source class injecting cosmic rays up to 1-100 EeV, with a luminosity-dependent injection power, i.e., $L_p\propto L_{100}$ \citep{Padovani_2015}. This ensures that more luminous sources contribute more to neutrino and IGRB backgrounds. If the protons are cooled sufficiently inside the jet, they produce neutrino fluxes depending on the target photon field luminosity due to $p\gamma$ interactions \citep{Murase_2014, Tavecchio_2014a, Palladino_2019}. On the contrary, we explicitly assume that the cosmic rays efficiently escape the system. This is justified as long as the escape rate is higher than the cooling rate of protons, and acceleration dominates up to the desired $E_{p,\rm max}$. 
Our model doesn't account for the sub-PeV neutrinos, which are expected to be dominated by neutrinos produced inside the high-energy sources. Similar results have been obtained using a likelihood analysis of the IceCube data in \cite{Kochoki_2020}, where they show interactions of cosmic rays from blazar AGNs can make up for $30-40\%$ of the diffuse flux. 

A strict $L_p/L_{100}$ correlation may not hold invariably for all sources. The ``blazar sequence'' predicts that BL Lacs with synchrotron and IC peak at higher energies are fainter in photon flux \citep{Fossati_1998, Ghisellini_2008, Ghisellini_2017}. The low-luminosity counterparts can be more hadronically powered (higher $\eta_{\rm eff}$) since the predicted SED peak energies are higher. Future observation of the $\gamma$-ray SED of such sources may provide further information. Here, we find that, for maximum proton acceleration energy $E_{p, \rm max}$ between 1 EeV and 100 EeV, the value of $\eta_{\rm eff}$ ranges from 11.1 to 4.4.  

The 4LAC catalog provides the redshift information for all the resolved FSRQs and most BL Lac objects. The deduced luminosity function of BL Lacs indicates that nearly $\sim65\%$ are yet to be identified. A majority of these are low-luminosity ($L_{\gamma}<10^{44}$ erg s$^{-1}$) counterparts and exhibit a negative redshift evolution. They are modeled as potential UHECR accelerators, owing to the preference of softer spectral index consistent with the Fermi acceleration model, compared to the hard injection required for a flat or positive redshift evolution \citep{Taylor_2015}. By including the unresolved sources, the neutrino flux increases by a factor of two at a few PeV. Thus, we can say that \textit{Fermi}-LAT has already detected a significant fraction of the AGNs that contributes to the neutrino flux at this energy, particularly the most luminous ones. 

Hadronic emission processes can also model the GeV-TeV $\gamma$-rays from BL Lacs and FSRQs \citep{Bottcher_2013, Petropoulou_2015, Sahu_2019}. Leptonic emission alone, too, can explain the SED up to TeV energies, with neutrinos originating from a radiatively subdominant hadronic component \citep{Keivani:2018rnh}. This is applicable to the 3$\sigma$ association of IceCube-170922A with the blazar TXS 0506+056, while alternate explanations involving $pp$ process \citep{Banik_2019} or multi-zone emission also exists \citep{Xue:2019txw, Xue:2020kuw}.  Recently, the IceCube-200107A event has been correlated with flaring blazar 3HSP J095507.9+355101 \citep{Giommi:2020viy} and  IC190730A  has been correlated with PKS 1502+106 \citep{Franckowiak:2020qrq}. However, a correlation of any neutrino event with nearby flaring blazars, such as MrK 421, Mrk 501 is not found. This can be due to a low neutrino flux level owing to lower cosmic-ray power. The deflection of 10 EeV UHECR protons from these sources at a distance of $\approx 150$ Mpc, in 0.1 nG magnetic field, can be $\sim1^\circ$ for a turbulence correlation length of 1 Mpc. Future detection of neutrinos and cosmic rays in spatial and temporal coincidence with nearby blazars will put our proposition concerning UHECR acceleration into firmer grounds.

The values of $E_{p, \rm max}$ considered in the study resembles the typical values obtained in the lepto-hadronic/hadronic modeling of blazar SEDs \citep{Mucke_2003, Bottcher_2013}. We see that beyond a maximum acceleration energy of $\approx 10$ EeV, the IGRB flux is saturated at TeV energies, thus requiring a lower baryon load, which in turn reduces the neutrino flux at a few PeV. Assuming a rigidity-dependent steepening of the cosmic ray spectrum, the knee for heavier primaries occur at $E\approx10^{16.92}$ eV \citep{Apel_2011, Apel_2013}, beyond which the proton abundance dominates at least up to $10^{18.2}$ eV. Our choice of injection spectral index $\alpha_p=2.6$ conforms with that considered in earlier studies for the extragalactic light component in this energy range \citep{Aloisio_2014, Liu_2016}.

Our analysis considers protons injected with a minimum energy of 10 PeV. In principle, protons of even lower energy can also escape, since we assume that the observed $\gamma$-rays from these blazars originate in leptonic processes only. This will increase the luminosity budget. However, there may also exist a break in the proton spectrum near to 10 PeV, preferring harder spectral index below ($\alpha_p<2$), thus reducing the luminosity requirement. Depending on the normalization and the total number of blazars obtained in more updated luminosity dependent density evolution functions \citep[see, eg.,][]{Qu_2019}, the estimates obtained here can change moderately. We do not explore the possible variation in our results due to different parameters of EM cascade such as the extragalactic magnetic fields, EBL models, etc. Earlier studies show that these effects are rather moderate \citep{Batista_2015}. However, we consider a discrete scenario, with an exemplary choice of parameters, that allows us to adequately explore cosmic-ray interactions on EBL in the context of diffuse PeV neutrino background. 

\section{Conclusions} \label{sec:conclusions}

We found that the resolved gamma-ray blazars from the \textit{Fermi}-4LAC catalog can explain up to $10\%$ of the IceCube diffuse neutrino flux upper limits at a few PeV energies. This requires a baryon load (cosmic-ray to gamma-ray luminosities) of $\approx 10$. FSRQs dominantly contribute to the neutrino flux. While including the unresolved gamma-ray blazars, the contribution can increase by a factor of two, depending on the maximum injected cosmic-ray energy. The baryon load in this case is bounded by the UHECR flux and varies between 4-11. The gamma-ray flux contribution, from the UHECR interactions, can be up to the Fermi diffuse flux upper limit at 820 GeV, depending upon the maximum injected cosmic-ray energy.



\software{\textsc{CRPropa 3} \citep{Batista_16}, DINT \citep{Lee_98, Heiter_18}}





\bibliography{icecube}{}
\bibliographystyle{aasjournal}



\end{document}